\RequirePackage{ifpdf}
\documentclass[12pt,letterpaper]{article}
\pdfoutput=1
\usepackage{jheppub}
\usepackage{epsfig}
\usepackage{enumitem}
\usepackage{bbm,amsfonts}
\usepackage{rotating,graphicx}
\usepackage{amssymb,amsmath,amsfonts,mathtools}
\usepackage{fancybox,diagbox}
\usepackage{enumerate}
\usepackage{dsfont}
\usepackage{verbatim}
\usepackage{wrapfig}
\usepackage{slashed}
\usepackage{braket}
\usepackage{pdflscape}
\usepackage{accents}
\usepackage{afterpage}
\usepackage{breqn}

\author[a]{Marco S. Bianchi}
 
\affiliation[a]{Facultad de Ingeniería, Arquitectura y Diseño, Universidad San Sebastián, Santiago, Chile}

\emailAdd{marco.bianchi@uss.cl}  
\abstract{We compute the two-point function of protected dimension-1 operators in ABJM up to two loops in dimensional regularization. The result exhibits uniform transcendentality empirically, which we conjecture to hold at all orders. We leverage this property to streamline the reconstruction of the dimensional regularization expansion of master integrals in terms of bases of Euler sums of uniform transcendental weight.}

\title{Transcendentality of ABJM two-point functions}


\def\Tr{\textrm{Tr}}

\numberwithin{equation}{section}

\begin{document}

\maketitle
\allowdisplaybreaks

\section{Introduction}

In this article, we investigate the transcendental structure of two-point functions in the ABJM model, the \(\mathcal{N}=6\) superconformal Chern-Simons theory in three spacetime dimensions \cite{Aharony:2008ug}. This is motivated by the results found in \(\mathcal{N}=4\) SYM, where uniform transcendentality appears in the perturbative computation of two-point functions of protected dimension-2 operators calculated in dimensional regularization \cite{Bianchi:2023llc}. We aim to determine whether and how a similar phenomenon occurs in the ABJM model.

To this purpose, we focus on two-point functions of protected scalar operators in the ABJM model of the form \( O = \Tr(A A) \), where \( A \) represents a complex scalar of the theory. While such correlators are tree-level exact in \(\mathcal{N}=4\) SYM (in exactly four dimensions), in the ABJM model only their dimension is protected from quantum corrections. Their normalization is not. 
We provide evidence that the dimensional regularization expansion of this normalization exhibits uniform transcendentality at two loops. We conjecture that this property extends to all orders.

Once established empirically, uniform transcendentality is then leveraged to facilitate an analytic evaluation of the non-trivial three-loop master integrals in momentum space appearing in the calculation. We perform high precision numerical evaluations of their expansions in dimensional regularization, via dimensional recurrence relations \cite{Lee:2009dh}. We then use the uniform transcendentality conjecture to construct combinations exhibiting this property. Finally, we  reconstruct their expansion coefficients as rational combinations of suitable bases of uniformly transcendental Euler sums via PSLQ \cite{Bailey:1991:PTN,Arno:1993:NPT,Bailey:1999nv} or LLL \cite{Lenstra:1982eee}. 

This paper is structured as follows.
In Section \ref{sec:2pt}, we provide a brief review of the relevant aspects of the ABJM model and introduce the scalar operators whose two-point functions we will compute.
In Section \ref{sec:calculation}, we perform the perturbative analysis of the two-point functions up to two-loop order.
In Section \ref{sec:ut}, we observe empirical evidence of uniform transcendentality and conjecture its extension to all loop orders.
In Section \ref{sec:constraints}, we expand the relevant master integrals to higher orders in dimensional regularization, assuming uniform transcendentality and corroborating its validity.
In Section \ref{sec:comparison} we  compare to $\mathcal{N}=4$ SYM in four dimensions.
Finally, we conclude with future perspectives.

\section{Two-point functions in the ABJM model}\label{sec:2pt}

We work in ABJM theory in three dimensions \cite{Aharony:2008ug}.
The coupling constant is identified with the inverse Chern-Simons level $k^{-1}$ and we will perform perturbation theory around $k\to\infty$. The other two parameters of the model are the ranks of the gauge groups, which we keep distinct: $N_1$ and $N_2$, as in \cite{Aharony:2008gk}.
We consider two-point functions of  dimension-1 protected operators in ABJM of the form
\begin{equation}
    \langle O(x) \bar O(0) \rangle \qquad O = \Tr(A A), \quad \bar O = \Tr(\bar A 
\bar A)
\end{equation}
with one scalar field $A$ of the theory. 
Their spacetime structure is fixed by conformal invariance in three dimensions
\begin{equation}
    \langle O(x) \bar O(0) = \frac{N(k)}{\left(x^2\right)^{\Delta}}
\end{equation}
Supersymmetry prevents the operator's dimension from renormalizing, so that it is fixed to 1
\begin{equation}
    \langle O(x) \bar O(0) = \frac{N(k)}{x^2}
\end{equation}
We consider its Fourier transform to momentum space, where the perturbative calculation we set out to undertake is more manageable. Then
\begin{equation}
    \langle O(p) \bar O(-p) = \frac{n(k)}{|p|}
\end{equation}
where we have modified the normalization accordingly and the spacetime dimension is strictly $d=3$.
Since the dependence on the momentum is fixed, we set $p^2=1$ throughout the rest of the paper and focus on the norm $n(k)$ in momentum space.
Unlike $\mathcal{N}=4$ SYM, such a numerator is not tree level exact and has been evaluated to two loops \cite{Young:2014lka,Bianchi:2020cfn} 
\begin{equation}
    n(k)= 2N_1N_2\left( \frac{1}{8}-\frac{\pi ^2 \left(N_1^2+N_2^2-2\right)}{48k^2} + \mathcal{O}\left( k^{-4} \right) \right)
\end{equation}
where the factor 2 emerges from the two identical contractions of the scalars and might be removed with a different operator normalization or choosing different field flavors.
We aim to study  the transcendental properties of $n(k)$, especially whether its perturbation theory hints at uniform transcendentality when expanded to higher orders in dimensional regularization $d=3-2\epsilon$.

The one loop correction to the two-point function vanishes identically. In fact, this extends to all odd loop orders. An odd number of antisymmetric Levi-Civita tensors appears at such perturbative orders, however only a single vector is present in the calculation: $x$ or $p$. Any contraction of indices with such a vector yields a vanishing result.
On the contrary, at even orders an even number of antisymmetric tensors appears, which evaluate in general to products of metric tensors.

For our analysis, vanishing of odd loop contributions is a nuisance, since the next non-trivial perturbative order is four loops. Such a calculation would involve a high computational complexity and entails the expansion to high orders in dimensional regularization of five-loop momentum integrals, which is currently unknown in three dimensions. 

Partial evidence for uniform transcendentality arises from the calculation of \cite{Bianchi:2016rub}, predicting that the maximally color imbalanced component of the two-point function is expressed to all orders, at $\epsilon=0$, by the expansion of 
\begin{equation}
    n(k) \Big|_{\text{maximal powers of } N_1} = 2N_1N_2\, \frac{k \sin \left(\frac{\pi  N_1}{k}\right)}{8 \pi  N_1}
\end{equation}
producing rational multiples of zetas $\zeta_L$ at even integer $L$ values, at loop $L$.

\section{The calculation}\label{sec:calculation}

The two-point function can be evaluated straightforwardly in terms of Feynman diagrams. A key subtlety is that the dimensional reduction scheme \cite{Siegel:1979wq} is essential for uniform transcendentality \cite{Bianchi:2013pva,Bianchi:2014iia}. This ensures proper handling of epsilon tensors and  $\gamma$ matrix algebra in the numerators.

We perform the calculation at two loops, separating the result according to the color factors
\begin{equation}\label{eq:n}
    n(k)= 2 N_1N_2 \left(n_0 +\frac{1}{k^2}\left( \left(N_1^2+N_2^2\right) c_{N_1^2}(\epsilon) + N_1N_2 c_{N_1N_2}(\epsilon) + c_1(\epsilon) \right) \right) + \mathcal{O}\left(k^{-4}\right)
\end{equation}
In momentum space, the tree level result is just the bubble integral
\begin{equation}
n_0 = \vcenter{\hbox{\includegraphics[scale=0.13]{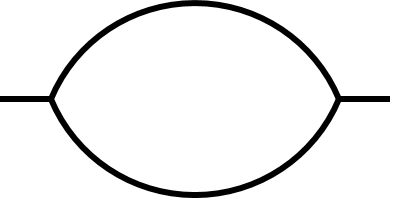}}} = G(1,1)
\end{equation}
where
\begin{equation}
    G(\alpha,\beta) \equiv  \frac{e^{\gamma  \epsilon } \Gamma \left(\frac{d}{2}-\alpha \right) \Gamma \left(\frac{d}{2}-\beta \right) \Gamma \left(\alpha +\beta-\frac{d}{2} \right)}{(4\pi)^{3/2} \Gamma (\alpha ) \Gamma (\beta ) \Gamma (d-\alpha -\beta )}
\end{equation}
By this normalization choice, we are discarding unimportant factors $e^{-\gamma \epsilon} (4\pi)^{\epsilon}$ for each momentum loop integration, which can be absorbed in the overall factor and in the coupling constant.
We recall that the external momentum scale $p^2$ was set to unity.

After the evaluation of Feynman diagrams, the result is reduced to master integrals. We used FIRE \cite{Smirnov:2008iw,Smirnov:2019qkx} and LiteRed \cite{Lee:2012cn,Lee:2013mka} for the task.
This leads to the following result
\begin{subequations}\label{eq:coeffs}
\begin{align}
\frac{c_{N_1^2}}{\pi^2}=&- \frac{16 (2d-5) (3d-8)}{(d-3)^2}\,\, \vcenter{\hbox{\includegraphics[scale=0.2]{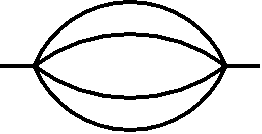}}}
+ \frac{16 (3 d-8)}{d-3}\,\, \vcenter{\hbox{\includegraphics[scale=0.2]{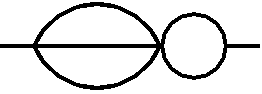}}}\nonumber\\&
- 32\,\, \vcenter{\hbox{\includegraphics[scale=0.2]{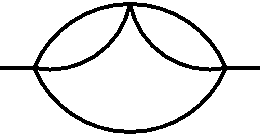}}} 
-16\,\, \vcenter{\hbox{\includegraphics[scale=0.2]{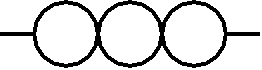}}}
+16\,\, \vcenter{\hbox{\includegraphics[scale=0.2]{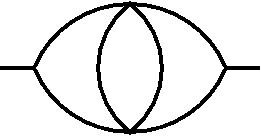}}} \label{eq:coeffs1}\\
\frac{c_{N_1N_2}}{\pi^2}&=32\,\,  \vcenter{\hbox{\includegraphics[scale=0.2]{MI4.png}}} + \frac{16 (3 d-8) ( 17 d^3 - 150 d^2+ 433 d + -406)}{(d-4)^2 (d-3) (2 d-7)}\,\, \vcenter{\hbox{\includegraphics[scale=0.2]{MI3.png}}} \nonumber\\&
- \frac{64 (11 d^2 - 64 d + 92)}{(d-4)^2}\,\, \vcenter{\hbox{\includegraphics[scale=0.2]{MI2.png}}}\nonumber\\&
+ \frac{32 (2 d-5) (3 d-8) (45 d^4 - 577 d^3 + 2775 d^2 - 5936 d+4768)}{(d-4)^3 (d-3)^2 (2 d-7)}\,\, \vcenter{\hbox{\includegraphics[scale=0.2]{MI1.png}}}\nonumber\\&
- \frac{8 (23 d^2 - 155 d+262)}{(d-4) (2 d-7)}\,\, \vcenter{\hbox{\includegraphics[scale=0.2]{MIP.png}}}
- \frac{4 (d-4)}{2 d-7}\,\,\vcenter{\hbox{\includegraphics[scale=0.2]{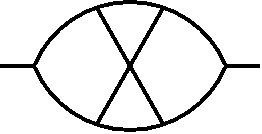}}} 
\label{eq:coeffs2}\\
\frac{c_{1}}{\pi^2}&=-\frac{112 (d-3) (3 d-10) (3 d-8)}{(d-4)^2 (2 d-7)}\,\, \vcenter{\hbox{\includegraphics[scale=0.2]{MI3.png}}}
+ \frac{768 (d-3)^2}{(d-4)^2}\,\, \vcenter{\hbox{\includegraphics[scale=0.2]{MI2.png}}} \nonumber\\&
 - \frac{32 (2 d-5) (3 d-8) ( 43 d^2- 288 d +480)}{(d-4)^3 (2 d-7)}\,\, \vcenter{\hbox{\includegraphics[scale=0.2]{MI1.png}}}\nonumber\\&
 + \frac{
 40 (d-3) (3 d-10)}{(d-4) (2 d-7)}\,\, \vcenter{\hbox{\includegraphics[scale=0.2]{MIP.png}}} + \frac{4 (d-4)}{(2 d-7)}\,\, \vcenter{\hbox{\includegraphics[scale=0.2]{MINP.png}}}\label{eq:coeffs3}
\end{align}
\end{subequations}
The diagrams above symbolize the respective master integrals. We have factored out a common $\pi^2$, for later convenience.

\section{Uniform transcendentality}\label{sec:ut}

The next step to inspect  transcendentality properties consists in expanding master integrals in dimensional regularization for $d=3-2\epsilon$. 
The simplest can be expressed in terms of $\Gamma$ functions and expanded straightforwardly
\begin{align}
\vcenter{\hbox{\includegraphics[scale=0.2]{MI1.png}}} &= G(1,1)G(1,2 \epsilon ) G\left(\frac{1}{2}+\epsilon,1\right)\nonumber\\
\vcenter{\hbox{\includegraphics[scale=0.2]{MI2.png}}} &= G(1,1)^2 G(1+2 \epsilon,1)\nonumber\\
\vcenter{\hbox{\includegraphics[scale=0.2]{MI3.png}}} &= G(1,1)^2 G\left(\frac{1}{2}+\epsilon,1\right)\nonumber\\
\vcenter{\hbox{\includegraphics[scale=0.2]{MI4.png}}} &= G(1,1)^3
\end{align}
The $\,\,\vcenter{\hbox{\includegraphics[scale=0.15]{MIP.png}}}\,\,$ master integral possesses an exact expression in terms of a hypergeometric function \cite{Kotikov:1995cw}
\begin{align}
\vcenter{\hbox{\includegraphics[scale=0.2]{MIP.png}}} &= 
\frac{e^{2\gamma\epsilon}}{(4\pi)^3}\, G(1,1)\, 2 \Gamma \left(5-\frac{3 d}{2}\right) \Gamma \left(\frac{d}{2}-1\right) \Gamma (d-3)
\\&
\left(\frac{\Gamma \left(\frac{d}{2}-1\right)}{(d-4) \Gamma \left(3-\frac{d}{2}\right) \Gamma (2 d-6)}\, 
_3F_2\left(\left.
\begin{array}{c}
1,4-d,d-2\\
5-d,3-\frac{d}{2}
\end{array}
\right|1\right)
-\frac{\pi  \cot \left(\frac{3 \pi  d}{2}\right)}{\Gamma (d-2)}\right)\nonumber
\end{align}
Its expansion around three dimensions can be performed with the algorithms of HypExp \cite{Huber:2005yg,Huber:2007dx} and HPL \cite{Maitre:2005uu,Maitre:2007kp}. However, at a certain order explicit expressions for harmonic polylogarithms at specific values are not tabled any longer. Hence, extracting analytic expressions a posteriori, as described momentarily, becomes more practical.
An expansion up to transcendental weight 6, i.e.~order $\epsilon^4$ can be found in \cite{Lee:2015eva} 
\begin{align}\label{eq:expMIP}
   &\vcenter{\hbox{\includegraphics[scale=0.2]{MIP.png}}} = 
   \frac{3 \zeta _2}{256}
   +\left(\frac{21 \zeta _3}{128}-\frac{3 L_1 \zeta _2}{128}\right) \epsilon \nonumber\\&~~~~
   +\left(\frac{L_1^4}{32}-\frac{21}{128} \zeta _2 L_1^2+\frac{21 \zeta _3 L_1}{64}+\frac{3L_4}{4}+\frac{669 \zeta _4}{1024}\right) \epsilon ^2\nonumber\\&~~~~
   +\left(\frac{3 L_1^5}{80}-\frac{9}{64} \zeta _2 L_1^3+\frac{21}{64} \zeta _3 L_1^2+\frac{3 L_4 L_1}{2}-\frac{669 \zeta _4L_1}{512}+3 L_5+\frac{75 \zeta _2 \zeta _3}{64}+\frac{1023 \zeta _5}{256}\right) \epsilon ^3\nonumber\\&~~~~
   +\left(\frac{7 L_1^6}{240}+\frac{31}{128} \zeta _2 L_1^4+\frac{7}{32} \zeta _3L_1^3+\frac{3}{2} L_4 L_1^2-\frac{2091}{512} \zeta _4 L_1^2+6 L_5 L_1+\frac{333}{64} \zeta _2 \zeta _3 L_1
   \right.\nonumber\\&~~~~\left.
   +\frac{1023 \zeta _5 L_1}{128}-\frac{129 \zeta _3^2}{16}+12 L_6+\frac{69L_4 \zeta _2}{8}+\frac{493773 \zeta _6}{16384}-\frac{33}{4} \zeta _{-5,-1}\right) \epsilon ^4+O\left(\epsilon ^5\right)
\end{align}
The following notation has been used for transcendental numbers in this formula
\begin{equation}
    L_n\equiv \text{Li}_n\left(\frac{1}{2}\right)
\end{equation}
so in particular $L_1 = \log 2$.
Multiple zeta values and Euler sums are defined according to 
\begin{equation}
    \zeta_{-5,-1} = -0.0299016\dots
\end{equation}
To the best of our knowledge, the non-planar master integral lacks a closed expression and has to be expanded with some suitable method.
The problem of propagator master integrals has been extensively analyzed in literature in four dimensions \cite{Lee:2011jt,Georgoudis:2018olj,Georgoudis:2021onj}.
Here we apply the dimensional recurrence relations method \cite{Lee:2009dh}, because the problem of expanding such master integrals has already been addressed, solved and coded in the heaven-sent package SummerTime \cite{Lee:2015eva} (see also \cite{Lee:2017ftw}).
With this implementation, expanding the integrals at higher orders in $\epsilon$ is fast and precise.
After suitably expanding numerically the integrals to some hundreds of digits, coefficients are reconstructed by Mathematica's implementations of the PSLQ and LLL algorithms.
For this task, a basis of transcendental numbers is necessary. Up to the orders we investigated in this work, it seems that Euler sums up to transcendental weight $l+2$ suffice for determining the expansions at order $\epsilon^l$ (though it was observed in \cite{Lee:2015eva} that this is not the case for four-loop integrals).
Up to order $\epsilon^4$, the non-planar master integral reads
\begin{align}\label{eq:expMINP}
\vcenter{\hbox{\includegraphics[scale=0.2]{MINP.png}}} =&~ \frac{\zeta _2}{16}-\frac{13}{64}+\left(\frac{\zeta _2 L_1}{2}-\frac{13 L_1}{32}-\frac{7 \zeta _2}{64}+\frac{17 \zeta _3}{32}-\frac{9}{8}\right) \epsilon \nonumber\\&
   +\left(-\frac{L_1^4}{4}+\frac{13}{8} \zeta _2 L_1^2
   -\frac{13 L_1^2}{32}+\frac{47 \zeta _2 L_1}{32}+\frac{17 \zeta _3 L_1}{16}-\frac{9 L_1}{4}-6 L_4\right.\nonumber\\&\left.-\frac{1319 \zeta_2}{128}-\frac{71 \zeta _3}{32}+\frac{19 \zeta _4}{2}+\frac{115}{16}\right) \epsilon ^2\nonumber\\&
   +\left(-\frac{3 L_1^5}{10}-\frac{9 L_1^4}{8}+\frac{4}{3} \zeta _2 L_1^3-\frac{13L_1^3}{48}+\frac{209}{32} \zeta _2 L_1^2+\frac{17}{16} \zeta _3 L_1^2\right.\nonumber\\&\left.-\frac{9 L_1^2}{4}-12 L_4 L_1-\frac{3599 \zeta _2 L_1}{64}
   -\frac{71 \zeta _3 L_1}{16}+\frac{1277 \zeta _4L_1}{32}+\frac{115 L_1}{8}-27 L_4\right.\nonumber\\&\left.
   -24 L_5+\frac{67 \zeta _2}{8}-\frac{249 \zeta _2 \zeta _3}{64}+\frac{307 \zeta _3}{64}+\frac{1519 \zeta _4}{256}+\frac{1637 \zeta _5}{32}-24\right)\epsilon ^3\nonumber\\&
   +\left(-\frac{7 L_1^6}{30}-\frac{27 L_1^5}{20}-\frac{17}{6} \zeta _2 L_1^4-\frac{37 L_1^4}{96}+\frac{263}{48} \zeta _2 L_1^3+\frac{17}{24} \zeta _3 L_1^3-\frac{3L_1^3}{2}
   \right.\nonumber\\&
   -12 L_4 L_1^2-\frac{10439}{64} \zeta _2 L_1^2-\frac{71}{16} \zeta _3 L_1^2+\frac{617}{8} \zeta _4 L_1^2+\frac{115 L_1^2}{8}-54 L_4 L_1
   \nonumber\\&
   -48 L_5 L_1+\frac{161 \zeta _2L_1}{2}-\frac{1653}{32} \zeta _2 \zeta _3 L_1+\frac{307 \zeta _3 L_1}{32}+\frac{13561 \zeta _4 L_1}{128}+\frac{1637 \zeta _5 L_1}{16}
   \nonumber\\&
   -48 L_1-\frac{1933 \zeta _3^2}{32}-6 L_4-108L_5-96 L_6-93 L_4 \zeta _2+\frac{1641 \zeta _2}{32}-\frac{999 \zeta _2 \zeta _3}{32}
   \nonumber\\&\left.
   +128 \zeta _3-\frac{618773 \zeta _4}{1024}+\frac{1433 \zeta _5}{64}+\frac{1689739 \zeta_6}{3072}+66 \zeta _{-5,-1}+\frac{235}{4}\right) \epsilon ^4+O\left(\epsilon ^5\right)
\end{align}
From the definition of the integrals or inspection into their expansion, it is clear that some simplifications are obtained by dividing by the one-loop bubble integral $\,\,\vcenter{\hbox{\includegraphics[scale=0.08]{M0L.png}}}\,\,$.
In terms of the two-point functions, this also serves the purpose of factoring the tree level result and normalizing the loop contributions by that.

Contrary to $\mathcal{N}=4$ SYM, such a normalization by the tree level result is not necessary for exposing uniform transcendentality. The tree level contribution is proportional to the integral $G(1,1)$, which happens to be uniformly transcendental in $d=3-2\epsilon$ dimensions. Hence, normalizing by this contribution does not alter the transcendentality properties of the two-point function. 

Plugging the integral expansions into \eqref{eq:coeffs} produces the following normalized two-loop corrections associated to different color structures
\begin{dmath}
  \frac{c_{N_1^2}}{n_0}=-\zeta _2+\epsilon  \left(19 \zeta _3-24 \zeta _2 L_1\right)+\epsilon ^2 \left(-30 \zeta _4-48 \zeta _2 L_1^2+4 L_1^4+96 L_4\right)+\epsilon ^3 \left(\frac{41 \zeta _2 \zeta
   _3}{3}+\frac{975 \zeta _5}{2}-32 \zeta _2 L_1^3-1002 \zeta _4 L_1-\frac{16 L_1^5}{5}+384 L_5\right)+\epsilon ^4 \left(-1056 \zeta _{-5,-1}-\frac{2675 \zeta _3^2}{3}-\frac{7075 \zeta
   _6}{16}-28 \zeta _2 L_1^4-1356 \zeta _4 L_1^2+568 \zeta _2 \zeta _3 L_1+864 \zeta _2 L_4+\frac{32 L_1^6}{15}+1536 L_6\right)+O\left(\epsilon ^5\right)
\end{dmath}
\begin{dmath}
\frac{c_{N_1N_2}}{n_0} =\epsilon  \left(36 \zeta _2 L_1-55 \zeta _3\right)+\epsilon ^2 \left(72 \zeta _2 L_1^2-\frac{463 \zeta _4}{2}\right)+\epsilon ^3 \left(139 \zeta _2 \zeta _3-2612 \zeta _5+96 \zeta _2
   L_1^3+564 \zeta _2^2 L_1\right)+\epsilon ^4 \left(\frac{11132 \zeta _3^2}{3}-\frac{369337 \zeta _6}{24}+128 \zeta _2 L_1^4+2340 \zeta _4 L_1^2+264 \zeta _2 \zeta _3 L_1+768 \zeta _2
   L_4\right)+O\left(\epsilon ^5\right)
\end{dmath}
\begin{dmath}
\frac{c_{1}}{n_0} = 2 \zeta _2+\epsilon  \left(17 \zeta _3+12 \zeta _2 L_1\right)+\epsilon ^2 \left(\frac{583 \zeta _4}{2}+24 \zeta _2 L_1^2-8 L_1^4-192 L_4\right)+\epsilon ^3 \left(-\frac{499 \zeta _2
   \zeta _3}{3}+1637 \zeta _5-32 \zeta _2 L_1^3+594 \zeta _4 L_1+\frac{32 L_1^5}{5}-768 L_5\right)+\epsilon ^4 \left(2112 \zeta _{-5,-1}-\frac{5782 \zeta _3^2}{3}+\frac{195281 \zeta
   _6}{12}-72 \zeta _2 L_1^4+372 \zeta _4 L_1^2-1400 \zeta _2 \zeta _3 L_1-2496 \zeta _2 L_4-\frac{64 L_1^6}{15}-3072 L_6\right)+O\left(\epsilon ^5\right)
\end{dmath}
The result is manifestly uniformly transcendental, up to the order to which it was expanded. A few more orders are indirectly evaluated below, where we leverage uniform transcendentality to streamline the expansion of the non-trivial master integrals.

The empirical evidence exposed above suggests that, analogously to $\mathcal{N}=4$ SYM in four dimensions, the two-point function of protected, dimension-1 operators in ABJM exhibits uniform transcendentality.
Unfortunately, the peculiarities of the ABJM perturbative expansion limit the scope of this evidence to only one non-trivial order, two loops.
What is more with respect to $\mathcal{N}=4$ SYM is that uniform transcendentality holds across the various color structures, providing different constraints on the transcendental structure of the master integrals.
Only two are independent though, since
\begin{equation}
    2c_{N_1^2} + c_{N_1N_2} + c_1 = 0
\end{equation}
to all orders in $\epsilon$, as ascertained at the level of master integrals \eqref{eq:coeffs}.
The choice $N_1=N_2=2$ yields the $\mathcal{N}=8$ BLG model \cite{Bagger:2007jr,Gustavsson:2007vu} which consequently also exhibits uniform transcendentality.

\section{Constraints on the transcendentality of master integrals}\label{sec:constraints}

In this section we discuss the consequences of the uniform transcendentality conjecture at the level of the master integrals. Since there are two independent combinations which exhibit the property, we can infer uniformly transcendental combinations involving the two non-trivial master integrals separately. 

In the coefficient $c_{N_1^2}$ \eqref{eq:coeffs1} we can observe that
$\vcenter{\hbox{\includegraphics[scale=0.15]{MI2.png}}}$ and $\vcenter{\hbox{\includegraphics[scale=0.15]{MI4.png}}}$ are independently uniformly transcendental, since they can be expressed in terms of $\Gamma$ functions of uniform transcendental expansions.
The other master integrals are made uniformly transcendental by the rescalings
$\left(1-4\epsilon\right)\left(1-6\epsilon\right)\,\, \vcenter{\hbox{\includegraphics[scale=0.15]{MI1.png}}}$ and $\left(1-6\epsilon\right)\,\,\vcenter{\hbox{\includegraphics[scale=0.15]{MI3.png}}}$.
Precisely the combination 
\begin{equation}
   \frac{4 (6 \epsilon -1)}{\epsilon ^2} \left((1-4 \epsilon)\,\,\vcenter{\hbox{\includegraphics[scale=0.2]{MI1.png}}}
   +2 \epsilon\,\, \vcenter{\hbox{\includegraphics[scale=0.2]{MI3.png}}}\right)
\end{equation}
appears in \eqref{eq:coeffs1}, upon setting $d=3-2\epsilon$.
Therefore the remaining integral in \eqref{eq:coeffs1} $\vcenter{\hbox{\includegraphics[scale=0.15]{MIP.png}}}$ has to be uniformly transcendental too, which is in fact the case. An explicit expansion is provided in \eqref{eq:expMIP} up to order $\epsilon^4$.
From a numerical expansion with SummerTime, we can reconstruct a few more expansion coefficients in terms of rational combinations of uniformly transcendental bases. 
The full bases of transcendental Euler sums with only an upper limit on transcendentality possess more elements. Hence, using the former demands less precision in the numerical evaluations and allow for much faster reconstructions. 
By taking the ratio with the one-loop bubble integral
\begin{equation}
    \bar V(\epsilon) \equiv \vcenter{\hbox{\includegraphics[scale=0.2]{MIP.png}}}\left/ \vcenter{\hbox{\includegraphics[scale=0.125]{M0L.png}}}\right.
\end{equation}
additional simplifications occur as various transcendentals constructed thanks to products of $\log 2$ dropping out of the expansion.
We state a couple further terms, up to transcendental weight 8 or $\epsilon^6$, to demonstrate this fact and corroborate the uniform transcendentality property
\begin{dmath}
   \bar V(\epsilon) = \frac{3 \zeta _2}{32}+\epsilon  \left(\frac{21 \zeta _3}{16}-\frac{3 \zeta _2 L_1}{8}\right)+\epsilon ^2 \left(\frac{297 \zeta _4}{64}-\frac{3}{4} \zeta _2 L_1^2+\frac{L_1^4}{4}+6
   L_4\right)+\epsilon ^3 \left(\frac{49 \zeta _2 \zeta _3}{8}+\frac{1023 \zeta _5}{32}+\zeta _2 L_1^3-\frac{297 \zeta _4 L_1}{16}-\frac{L_1^5}{5}+24 L_5\right)+\epsilon ^4 \left(-66
   \zeta _{-5,-1}-\frac{1025 \zeta _3^2}{16}+\frac{111783 \zeta _6}{512}+\frac{5}{4} \zeta _2 L_1^4+\frac{27}{8} \zeta _4 L_1^2+\frac{91}{4} \zeta _2 \zeta _3 L_1+54 \zeta _2
   L_4+\frac{2 L_1^6}{15}+96 L_6\right)+\epsilon ^5 \left(-\frac{960}{7} \zeta _{-5,1,1}+\frac{888}{7} \zeta _{5,-1,-1}+\frac{960}{7} L_1 \zeta _{-5,-1}+\frac{283419 \zeta _3 \zeta
   _4}{448}+\frac{105729 \zeta _2 \zeta _5}{160}+\frac{375 \zeta _7}{2}-\zeta _2 L_1^5-\frac{56}{3} \zeta _3 L_1^4-\frac{9}{2} \zeta _4 L_1^3+\frac{133}{2} \zeta _2 \zeta _3
   L_1^2-\frac{1023}{4} \zeta _5 L_1^2-\frac{1200}{7} \zeta _3^2 L_1-\frac{585345 \zeta _6 L_1}{896}+216 \zeta _2 L_5-448 \zeta _3 L_4-\frac{8 L_1^7}{105}+384 L_7\right)\\
   +\epsilon ^6
   \left(\frac{25440}{7} \zeta _{-7,-1}-\frac{15534}{7} \zeta _2 \zeta _{-5,-1}+\frac{158559 \zeta _{5,3}}{560}-\frac{3552}{7} \zeta _{-5,-1,-1,-1}-\frac{3840}{7} \zeta
   _{-5,-1,1,1}+\frac{1776}{7} L_1^2 \zeta _{-5,-1}-\frac{166843}{168} \zeta _2 \zeta _3^2-\frac{648799 \zeta _3 \zeta _5}{560}+\frac{199150479 \zeta _8}{17920}+\frac{2}{3} \zeta _2
   L_1^6+\frac{608}{105} \zeta _3 L_1^5+\frac{2747}{16} \zeta _4 L_1^4+\frac{58}{21} \zeta _2 \zeta _3 L_1^3+\frac{999}{14} \zeta _3^2 L_1^2-\frac{490401}{448} \zeta _6
   L_1^2-\frac{4869}{56} \zeta _3 \zeta _4 L_1-\frac{134349}{140} \zeta _2 \zeta _5 L_1+864 \zeta _2 L_6-\frac{4864 \zeta _3 L_5}{7}+\frac{8025 \zeta _4 L_4}{2}+\frac{4
   L_1^8}{105}+1536 L_8\right) + \mathcal{O}\left(\epsilon^7\right)
\end{dmath}
Two additional terms, up transcendental weight 10, are displayed explicitly in the appendix, since they start becoming bulky.
Different choices of basis elements for Euler sums are possible.

From relation \eqref{eq:coeffs3} we can infer that the lower transcendental part of the non-planar master integral is exactly provided, order-by-order, by the negative of the lower transcendental terms of all other contributions
\begin{align}
   & \vcenter{\hbox{\includegraphics[scale=0.2]{MINP.png}}}\,\,\bigg|_{\text{lower transc}}=
    \bigg(\frac{8 (1-4 \epsilon) (1-6 \epsilon) \left(172 \epsilon ^2+60 \epsilon +3\right)}{(1+2 \epsilon)^4}\,\,\vcenter{\hbox{\includegraphics[scale=0.2]{MI1.png}}}+\nonumber\\&
    \frac{56(1-6 \epsilon) (1+6 \epsilon) \epsilon }{(1+2 \epsilon)^3}\,\,\vcenter{\hbox{\includegraphics[scale=0.2]{MI3.png}}}-\frac{768 (1+4 \epsilon) \epsilon ^2}{(1+2 \epsilon)^3}\,\,\vcenter{\hbox{\includegraphics[scale=0.2]{MI2.png}}}\nonumber\\&
   -\frac{20 (1+6 \epsilon) \epsilon }{(1+2 \epsilon)^2}\,\, \vcenter{\hbox{\includegraphics[scale=0.2]{MIP.png}}}\,\,\bigg|_{\text{lower transc}}
\end{align}
Surprisingly, the terms on the right-hand-side of this equation turn out to provide exactly the lower transcendental part of the non-planar integral 
\begin{align}
   & \vcenter{\hbox{\includegraphics[scale=0.2]{MINP.png}}}\,\,\bigg|_{\text{lower transc}}=
    \frac{8 (1-4 \epsilon) (1-6 \epsilon) \left(172 \epsilon ^2+60 \epsilon +3\right)}{(1+2 \epsilon)^4}\,\,\vcenter{\hbox{\includegraphics[scale=0.2]{MI1.png}}}+\\&
    \frac{56(1-6 \epsilon) (1+6 \epsilon) \epsilon }{(1+2 \epsilon)^3}\,\,\vcenter{\hbox{\includegraphics[scale=0.2]{MI3.png}}}-\frac{768 (1+4 \epsilon) \epsilon ^2}{(1+2 \epsilon)^3}\,\,\vcenter{\hbox{\includegraphics[scale=0.2]{MI2.png}}} 
   -\frac{20 (1+6 \epsilon) \epsilon }{(1+2 \epsilon)^2}\,\, \vcenter{\hbox{\includegraphics[scale=0.2]{MIP.png}}}\nonumber
\end{align}
This is an empirical observation and we lack an explanation for it.
As a result, the expansion of $c_1$ in \eqref{eq:coeffs3} coincides with the maximally transcendental part of the non-planar master integral.
Conversely, the expression
\begin{align}\label{eq:MINPut}
   & U(\epsilon)\equiv\vcenter{\hbox{\includegraphics[scale=0.2]{MINP.png}}}-
    \frac{8 (1-4 \epsilon) (1-6 \epsilon) \left(172 \epsilon ^2+60 \epsilon +3\right)}{(1+2 \epsilon)^4}\,\,\vcenter{\hbox{\includegraphics[scale=0.2]{MI1.png}}}\\&
    -\frac{56(1-6 \epsilon) (1+6 \epsilon) \epsilon }{(1+2 \epsilon)^3}\,\,\vcenter{\hbox{\includegraphics[scale=0.2]{MI3.png}}}+\frac{768 (1+4 \epsilon) \epsilon ^2}{(1+2 \epsilon)^3}\,\,\vcenter{\hbox{\includegraphics[scale=0.2]{MI2.png}}} 
   +\frac{20 (1+6 \epsilon) \epsilon }{(1+2 \epsilon)^2}\,\, \vcenter{\hbox{\includegraphics[scale=0.2]{MIP.png}}}\nonumber
\end{align}
is, conjecturally, uniformly transcendental to all orders in $\epsilon$. 

We leverage this conjecture to facilitate the analytic extraction of its coefficients.
A numeric evaluation of \eqref{eq:MINPut} can be fitted to numbers belonging to a uniform transcendental basis.
The working hypothesis for this integral is that Euler sums are sufficient for the task.
A few simplifications occur after normalizing by the one-loop bubble integral
\begin{equation}
    \bar U(\epsilon) \equiv U(\epsilon) \left/\vcenter{\hbox{\includegraphics[scale=0.13]{M0L.png}}}\right.
\end{equation}
The expansion of such a combination reads, up to transcendental weight 8
\begin{dmath}
   \bar U(\epsilon) = 
2 \zeta _2+\epsilon  \left(17 \zeta _3+12 \zeta _2 L_1\right)+\epsilon ^2 \left(\frac{583 \zeta _4}{2}+24 \zeta _2 L_1^2-8 L_1^4-192 L_4\right)+\epsilon ^3 \left(-\frac{499 \zeta _2
   \zeta _3}{3}+1637 \zeta _5-32 \zeta _2 L_1^3+594 \zeta _4 L_1+\frac{32 L_1^5}{5}-768 L_5\right)+\epsilon ^4 \left(2112 \zeta _{-5,-1}-\frac{5782 \zeta _3^2}{3}+\frac{195281 \zeta
   _6}{12}-72 \zeta _2 L_1^4+372 \zeta _4 L_1^2-1400 \zeta _2 \zeta _3 L_1-2496 \zeta _2 L_4-\frac{64 L_1^6}{15}-3072 L_6\right)
   \\+\epsilon ^5 \left(\frac{73728}{7} \zeta
   _{-5,1,1}+\frac{14592}{7} \zeta _{5,-1,-1}-\frac{73728}{7} L_1 \zeta _{-5,-1}-\frac{3648733 \zeta _3 \zeta _4}{84}+\frac{69111 \zeta _2 \zeta _5}{5}+57908 \zeta _7-\frac{1376}{5}
   \zeta _2 L_1^5+\frac{3136}{3} \zeta _3 L_1^4+3984 \zeta _4 L_1^3-8848 \zeta _2 \zeta _3 L_1^2+8184 \zeta _5 L_1^2+\frac{92160}{7} \zeta _3^2 L_1-9216 \zeta _2 L_4 L_1+\frac{883713
   \zeta _6 L_1}{28}-16128 \zeta _2 L_5+25088 \zeta _3 L_4+\frac{256 L_1^7}{105}-12288 L_7\right)\\
   +\epsilon ^6 \left(-\frac{814080}{7} \zeta _{-7,-1}+\frac{394944}{7} \zeta _2 \zeta
   _{-5,-1}+\frac{26946 \zeta _{5,3}}{35}+\frac{113664}{7} \zeta _{-5,-1,-1,-1}+\frac{122880}{7} \zeta _{-5,-1,1,1}-\frac{56832}{7} L_1^2 \zeta _{-5,-1}+\frac{706528}{63} \zeta _2
   \zeta _3^2-\frac{17904254 \zeta _3 \zeta _5}{105}+\frac{2980042927 \zeta _8}{5040}-704 \zeta _2 L_1^6+\frac{18176}{105} \zeta _3 L_1^5-1510 \zeta _4 L_1^4-\frac{114752}{21} \zeta _2
   \zeta _3 L_1^3-\frac{15984}{7} \zeta _3^2 L_1^2-18432 \zeta _2 L_4 L_1^2+\frac{751473}{14} \zeta _6 L_1^2-12288 \zeta _2 L_5 L_1-\frac{241596}{7} \zeta _3 \zeta _4
   L_1+\frac{2064312}{35} \zeta _2 \zeta _5 L_1-39936 \zeta _2 L_6-\frac{145408 \zeta _3 L_5}{7}-171024 \zeta _4 L_4+\frac{4352 L_1^8}{105}+2048 L_4 L_1^4+24576 L_4^2-49152
   L_8\right) + \mathcal{O}\left(\epsilon^7\right)
\end{dmath}
and two additional orders are presented in the appendix.
The fact that these expansions can be determined in terms of uniformly transcendental numbers is a corroboration of the uniform transcendentality conjecture.

From a practical perspective, working with uniformly transcendental objects offers substantial advantages in the reconstruction.
For instance, fixing an analytic form for $U(\epsilon)$ at order $\epsilon^5$ and $\epsilon^6$ (transcendental weights 7 and 8) 
can be performed with basis sizes of 21 and 34 independent Euler sums.
The reconstruction achieves stability around order 180 and 380 digits, respectively.
Full bases of all independent Euler sums up to transcendental weights 7 and 8 have sizes 54 and 88. In those cases, an effective coefficient reconstruction of the non-planar master integral would require order 570 and 1250 digits, respectively.

Since the number of linearly independent basis elements for Euler sums at fixed transcendental weight $n$ grows according to the Fibonacci sequence $F_{n+1}$ \cite{Blumlein:2009cf}, the number of elements in a full basis considering transcendentality $\leq l$ is $$\sum_{n=0}^l F_{n+1} = \frac{1}{\sqrt{5}} \left(\left(-\left(2+\sqrt{5}\right)\phi\right)^{-l} + \left(2+\sqrt{5}\right)\phi^l\right) -1$$
where $\phi$ is the golden ratio.
The full basis is asymptotically $\phi^2$ times larger than the uniformly transcendental one, with a relative difference asymptoting $\phi$.

\section{Comparison to four dimensions}\label{sec:comparison}
An analogous observation of uniform transcendentality for two-point functions of lowest dimensions protected operators in $\mathcal{N}=4$ SYM was put forward in \cite{Bianchi:2023llc}.
Conjecturing it holds to all orders in $\epsilon$, it implies that the following combination of three-loop master integrals is uniformly transcendental when expanded at $d=4-2\epsilon$
\begin{align}
\frac{1}{G(1,1)}&\Bigg(\vcenter{\hbox{\includegraphics[scale=0.2]{MINP.png}}}
+\frac{16 (d-3)^2}{(d-4)^2} \vcenter{\hbox{\includegraphics[scale=0.2]{MI4.png}}}
\nonumber\\&
+\frac{16 (2 d-5) (3 d-8) (d (9 d-65)+118)}{(d-4)^4} \vcenter{\hbox{\includegraphics[scale=0.2]{MI1.png}}}
\nonumber\\&
+\frac{128 (2 d-7) (d-3)^2}{(d-4)^3} \vcenter{\hbox{\includegraphics[scale=0.2]{MI2.png}}}-\frac{(48 (3 d-10) (3 d-8) (d-3)}{(d-4)^3} \vcenter{\hbox{\includegraphics[scale=0.2]{MI3.png}}}
\nonumber\\&
+\frac{(12 (3 d-10) (d-3)}{(d-4)^2} \vcenter{\hbox{\includegraphics[scale=0.2]{MIP.png}}}\Bigg)
\end{align}
This observation can then be used to facilitate reconstructing the expansion coefficients of the non-planar master from numerics at high order in $\epsilon$, if needed.

At four loops the question of uniformly transcendental master integrals is more interesting.
The uniform transcendentality conjecture for two-point functions of dimension-2 operators only provides one constraint, over order twenty master integrals, which is not sufficient for extracting useful information. In case other independent constraints can be derived from other observables, then the situation could improve.

\section{Conclusions}

In this work we conjecture that the dimensional regularization expansion of the two-point function of supersymmetric dimension-1 operators in ABJM exhibits uniform transcendentality.
This is an empirical perturbative statement verified only at two-loop order and up to a certain fixed power of the regulator $\epsilon$, that is $\epsilon^{8}$. The extension to further perturbative orders, and to the whole $\epsilon$ expansion can only be conjectured.
As additional support, the same phenomenon seems to occur in four dimensions for $\mathcal{N}=4$ SYM theory \cite{Bianchi:2023llc}. In that context, a few more data points are accessible, since it is possible to extract non-trivial results at odd loop orders, where the ABJM analogues vanish trivially. 

While in $\mathcal{N}=4$ SYM no transcendentals outside the MZV realm pop up at three-loop perturbative order (four loop momentum integrals), the uniform transcendentality conjecture in ABJM involves Euler sums already at two-loop order. This causes a more demanding extraction of analytic coefficients from numerics, since the basis of numbers whose coefficients are unknown is larger. This in turn requires additional precision in the numerical evaluation. We have leveraged the uniform transcendentality conjecture to construct a combination of master integrals of uniform transcendental weight, whose analytic determination is then more straightforward.
We have estimated the relative advantage of a uniformly transcendental basis of Euler sums over the complete one, just because its asymptotics are governed by the golden ratio, henceforth aesthetically satisfactory. 

If a similar uniform transcendentality statement could be put forward for some two-point correlator in ABJM requiring four-loop master integrals in momentum space, it would likely involve transcendentals beyond Euler sums, whose appearance was diagnosed in \cite{Lee:2015eva}. This would constitute a natural development of the present work.

In \cite{Bianchi:2023llc} a relation was observed between the two-point function of protected operators and their three-point function in the limit of a soft external momentum. It would be interesting to explore whether a similar relation holds for three-point functions in ABJM. This might shed additional light on conflicting and not completely satisfactory results in literature for such three-point functions \cite{Young:2014lka,Bianchi:2020cfn}.

\acknowledgments

This work was supported by Fondo Nacional de Desarrollo Cient\'ifico y Tecnol\'ogico, through Fondecyt Regular 1220240 and Fondecyt Exploraci\'on 13220060.

\vfill
\newpage

\appendix

\section{Expansions up to transcendental weight 10}
\subsection{Planar non-trivial master integral}
At transcendental weight 9 $\bar V(\epsilon)$ reads
\begin{align}
\bar V(\epsilon)^{(7)} =& ~\frac{285092}{91} \zeta _3 \zeta _{-5,-1}-528 \zeta _{-7,1,1}-\frac{290088}{91} \zeta _2 \zeta _{-5,1,1}-\frac{10992}{13} \zeta _2
   \zeta _{5,-1,-1}\nonumber\\&-\frac{344112}{91} \zeta _{7,-1,-1}+\frac{1824}{91} \zeta _{-5,-1,-1,-1,1}+\frac{94176}{91} \zeta _{-5,-1,-1,1,1}\nonumber\\&-\frac{31392}{91} \zeta
   _{-5,-1,1,-1,-1}-\frac{8256}{7} \zeta _{-5,-1,1,1,1}+\frac{94480}{91} L_1^3 \zeta _{-5,-1}+\frac{15696}{91} L_1^2 \zeta _{-5,1,1}\nonumber\\&+\frac{78480}{91} L_1^2 \zeta
   _{5,-1,-1}+\frac{1319424}{91} L_1 \zeta _{-7,-1}+\frac{383352}{91} \zeta _2 L_1 \zeta _{-5,-1}+\frac{450279}{280} L_1 \zeta _{5,3}\nonumber\\&-\frac{188352}{91} L_1 \zeta
   _{-5,-1,-1,-1}+\frac{94176}{91} L_1 \zeta _{-5,-1,1,1}+\frac{7078265 \zeta _3^3}{4368}-\frac{227534193 \zeta _4 \zeta _5}{58240}\nonumber\\&+\frac{12805099 \zeta _3 \zeta
   _6}{2912}+\frac{9236757 \zeta _2 \zeta _7}{2912}+\frac{1156507903 \zeta _9}{11648}+\frac{9944 \zeta _2 L_1^7}{1365}+\frac{3313}{819} \zeta _3 L_1^6\nonumber\\&-\frac{82595}{364} \zeta _4
   L_1^5-\frac{93097}{273} \zeta _2 \zeta _3 L_1^4-\frac{362133}{455} \zeta _5 L_1^4+\frac{50607}{91} \zeta _3^2 L_1^3+\frac{10464}{91} \zeta _2 L_4 L_1^3\nonumber\\&-\frac{5798693 \zeta _6
   L_1^3}{1456}-\frac{31392}{91} \zeta _2 L_5 L_1^2+\frac{7848}{13} \zeta _3 L_4 L_1^2-\frac{92811}{52} \zeta _3 \zeta _4 L_1^2\nonumber\\&-\frac{9117039 \zeta _2 \zeta _5
   L_1^2}{1820}+\frac{412767}{364} \zeta _7 L_1^2-\frac{6308713 \zeta _2 \zeta _3^2 L_1}{1092}+\frac{141264}{91} \zeta _3 L_5 L_1\nonumber\\&-\frac{19620}{7} \zeta _4 L_4 L_1+\frac{2023293}{364}
   \zeta _3 \zeta _5 L_1-\frac{3070273579 \zeta _8 L_1}{58240}+3456 \zeta _2 L_7\nonumber\\&-\frac{535456 \zeta _3 L_6}{91}-\frac{832920}{91} \zeta _2 \zeta _3 L_4+\frac{378438 \zeta _4
   L_5}{91}-\frac{198936 \zeta _5 L_4}{65}-\frac{6094 L_1^9}{12285}\nonumber\\&-\frac{5232}{455} L_4 L_1^5+\frac{5232}{91} L_5 L_1^4+\frac{125568 L_4 L_5}{91}+6144
   L_9
\end{align}
and at transcendental weight 10
\begin{align}
   &\bar V(\epsilon)^{(8)} = -\frac{9535368644 L_1^{10}}{2204604675}+\frac{10136039869 \zeta _2 L_1^8}{146973645}-\frac{67459049044 \zeta _3 L_1^7}{1028815515}\nonumber\\&~~
   -\frac{5155105472 L_4L_1^6}{48991215}-\frac{137185298171 \zeta _4 L_1^6}{97982430}+\frac{718840064 L_5 L_1^5}{1399749}\nonumber\\&~~-\frac{657074324446 \zeta _2 \zeta _3 L_1^5}{342938505}+\frac{2154785922977 \zeta _5
   L_1^5}{489912150}+\frac{366988482535 \zeta _3^2 L_1^4}{117578916}\nonumber\\&~~-\frac{133506303839 \zeta _6 L_1^4}{29861312}+\frac{17736241944 \zeta
   _{-5,-1} L_1^4}{3266081}-\frac{521024256 L_5 \zeta _2 L_1^3}{171899}\nonumber\\&~~+\frac{371882464 L_4 \zeta _3 L_1^3}{466583}-\frac{29277789236 \zeta _3 \zeta _4
   L_1^3}{22862567}-\frac{1295291888057 \zeta _2 \zeta _5 L_1^3}{16330405}\nonumber\\&~~+\frac{307275852353 \zeta _7 L_1^3}{3266081}+\frac{2831273728 \zeta _{-5,1,1} L_1^3}{3266081}+\frac{7514040064
   \zeta _{5,-1,-1} L_1^3}{3266081}\nonumber\\&~~-\frac{3594946629991 \zeta _2 \zeta _3^2 L_1^2}{68587701}+\frac{5111506944 L_5 \zeta _3 L_1^2}{466583}-\frac{6635190808 L_4 \zeta _4
   L_1^2}{251237}\nonumber\\&~~+\frac{1426412816737 \zeta _3 \zeta _5 L_1^2}{6532162}-\frac{16459147570365 \zeta _8 L_1^2}{77010752}+\frac{57941015760 \zeta _{-7,-1}
   L_1^2}{3266081}\nonumber\\&~~+\frac{1120484738784 \zeta _2 \zeta _{-5,-1} L_1^2}{22862567}+\frac{1968005646 \zeta _{5,3} L_1^2}{251237}-\frac{32928721920 \zeta _{-5,-1,-1,-1}
   L_1^2}{3266081}\nonumber\\&~~+\frac{10156179664 L_4 \zeta _2 L_1^4}{9798243}+\frac{1560336768 \zeta _{-5,-1,1,1} L_1^2}{466583}+\frac{1156584567818 \zeta _3^3 L_1}{68587701}\nonumber\\&~~+\frac{41240843776 L_4 L_5 L_1}{3266081}-\frac{36664186752 L_6 \zeta
   _3 L_1}{3266081}-\frac{215749225920 L_4 \zeta _2 \zeta _3 L_1}{3266081}\nonumber\\&~~-\frac{113959525872 L_5 \zeta _4 L_1}{3266081}+\frac{67433321120 L_4 \zeta _5
   L_1}{466583}-\frac{180965609690483 \zeta _4 \zeta _5 L_1}{457251340}\nonumber\\&~~-\frac{199201020799687 \zeta _3 \zeta _6 L_1}{1463204288}+\frac{80779677613 \zeta _2 \zeta _7
   L_1}{6532162}+\frac{132095610426391 \zeta _9 L_1}{313543776}\nonumber\\&~~-\frac{872653386616 \zeta _3 \zeta _{-5,-1} L_1}{22862567}+\frac{2288880480 \zeta _{-7,1,1}
   L_1}{35891}-\frac{110366345088 \zeta _2 \zeta _{-5,1,1} L_1}{3266081}\nonumber\\&~~-\frac{100802498752 \zeta _2 \zeta _{5,-1,-1} L_1}{3266081}-\frac{13974354528 \zeta _{7,-1,-1}
   L_1}{3266081}+\frac{2411892096 \zeta _{-5,-1,-1,-1,1} L_1}{466583}\nonumber\\&~~+\frac{16464360960 \zeta _{-5,-1,-1,1,1} L_1}{3266081}+\frac{2923810688 \zeta _{-5,-1,1,-1,-1}
   L_1}{3266081}\nonumber\\&~~-\frac{2575785984 \zeta _{-5,-1,1,1,1} L_1}{251237}+\frac{189576960 L_5^2}{251237}+\frac{1994713814 L_4 \zeta _3^2}{753711}-\frac{4353274155729 \zeta
   _5^2}{281385440}\nonumber\\&~~+24576 L_{10}-\frac{47394240 L_4^2 \zeta _2}{251237}+13824 L_8 \zeta _2-\frac{285952 L_7 \zeta _3}{7}-\frac{2219497680 L_5 \zeta _2 \zeta
   _3}{1758659}\nonumber\\&~~-\frac{3887701666323 \zeta _3^2 \zeta _4}{112554176}+\frac{9353977032 L_6 \zeta _4}{251237}+\frac{46679978124 L_5 \zeta _5}{1256185}\nonumber\\&~~-\frac{7927859205619 \zeta _2 \zeta
   _3 \zeta _5}{140692720}+\frac{467151029397 L_4 \zeta _6}{2009896}-\frac{452601269 \zeta _3 \zeta _7}{502474}\nonumber\\&~~+\frac{36056747508643599 \zeta _{10}}{72034672640}+\frac{95686736352
   \zeta _{-9,-1}}{251237}-\frac{96851640888 \zeta _2 \zeta _{-7,-1}}{1758659}\nonumber\\&~~-\frac{362346048 L_4 \zeta _{-5,-1}}{35891}-\frac{556597338369 \zeta _4 \zeta
   _{-5,-1}}{3517318}+\frac{99612869469 \zeta _2 \zeta _{5,3}}{28138544}\nonumber\\&~~+\frac{698630689767 \zeta _{7,3}}{112554176}-\frac{23010936400 \zeta _3 \zeta
   _{-5,1,1}}{1758659}+\frac{136155400 \zeta _3 \zeta _{5,-1,-1}}{1758659}\nonumber\\&~~-\frac{4631578080 \zeta _{-7,-1,-1,-1}}{251237}-\frac{7269041952 \zeta _{-7,-1,1,1}}{251237}-\frac{6435234432
   \zeta _2 \zeta _{-5,-1,-1,-1}}{1758659}\nonumber\\&~~-\frac{17006670432 \zeta _2 \zeta _{-5,-1,1,1}}{1758659}-\frac{1018001664 \zeta _{-5,1,-1,-1,-1,-1}}{251237}\nonumber\\&~~-\frac{94788480 \zeta
   _{-5,1,1,-1,1,-1}}{251237}+\frac{798079872 \zeta _{-5,1,1,1,-1,-1}}{251237}-\frac{852615936 \zeta _{-5,1,1,1,1,1}}{251237}
\end{align}
The rational coefficients experience a suspicious jump upwards in complexity starting at transcendental weight 10, but this is due to the presence of a large ubiquitous prime number, 1889. A similar phenomenon occurs for MZV's at weight 12, due to the large prime 691 in the numerator of the Bernoulli number $B_{12}$.
The stability of the reconstruction of the coefficients has been tested across several hundreds of precision in the numerical evaluation of the integrals, beyond its onset.

\subsection{Non-planar uniformly transcendental combination of master integrals}

The expansion coefficients of $\bar U(\epsilon)$ up to transcendental weight 10 read 
\begin{dmath}   
\bar U(\epsilon)^{(7)} =  \frac{11255680}{91} \zeta _3 \zeta _{-5,-1}-62976 \zeta _{-7,1,1}+\frac{32172288}{91} \zeta _2 \zeta _{-5,1,1}-\frac{4875776}{91} \zeta _2 \zeta
   _{5,-1,-1}-\frac{12255744}{91} \zeta _{7,-1,-1}+\frac{1747968}{91} \zeta _{-5,-1,-1,-1,1}+\frac{3265536}{91} \zeta _{-5,-1,-1,1,1}+\frac{1893376}{91} \zeta
   _{-5,-1,1,-1,-1}+\frac{608256}{7} \zeta _{-5,-1,1,1,1}+\frac{5047808}{91} L_1^3 \zeta _{-5,-1}-\frac{946688}{91} L_1^2 \zeta _{-5,1,1}-\frac{260608}{91} L_1^2 \zeta
   _{5,-1,-1}+\frac{49815552}{91} L_1 \zeta _{-7,-1}-\frac{25587456}{91} \zeta _2 L_1 \zeta _{-5,-1}+\frac{1951668}{35} L_1 \zeta _{5,3}-\frac{6531072}{91} L_1 \zeta
   _{-5,-1,-1,-1}+\frac{3265536}{91} L_1 \zeta _{-5,-1,1,1}+\frac{87515314 \zeta _3^3}{819}-\frac{1479701267 \zeta _4 \zeta _5}{1820}-\frac{40907276245 \zeta _3 \zeta
   _6}{13104}-\frac{3366953 \zeta _2 \zeta _7}{91}+\frac{23121935561 \zeta _9}{3276}+\frac{233216 \zeta _2 L_1^7}{4095}-\frac{8295712 \zeta _3 L_1^6}{4095}-\frac{6506952}{455} \zeta _4
   L_1^5+\frac{3811552}{91} \zeta _2 \zeta _3 L_1^4-\frac{43629632 \zeta _5 L_1^4}{1365}-\frac{29920}{91} \zeta _3^2 L_1^3-\frac{4129792}{273} \zeta _2 L_4 L_1^3-\frac{4953750}{91}
   \zeta _6 L_1^3-\frac{9288704}{91} \zeta _2 L_5 L_1^2-\frac{473344}{13} \zeta _3 L_4 L_1^2-\frac{11126216}{13} \zeta _3 \zeta _4 L_1^2+\frac{113917112}{455} \zeta _2 \zeta _5
   L_1^2+\frac{39982056}{91} \zeta _7 L_1^2+\frac{102468008}{273} \zeta _2 \zeta _3^2 L_1-147456 \zeta _2 L_6 L_1+\frac{4898304}{91} \zeta _3 L_5 L_1-\frac{4547456}{7} \zeta _4 L_4
   L_1+\frac{6615288}{91} \zeta _3 \zeta _5 L_1-\frac{587857077 \zeta _8 L_1}{1820}-258048 \zeta _2 L_7-\frac{3939328 \zeta _3 L_6}{91}+\frac{67331840}{91} \zeta _2 \zeta _3
   L_4-\frac{135553728 \zeta _4 L_5}{91}-\frac{38468608 \zeta _5 L_4}{65}-\frac{476992 L_1^9}{12285}-\frac{1289728 L_4 L_1^5}{1365}+\frac{1289728}{273} L_5 L_1^4+\frac{10317824 L_4
   L_5}{91}-196608 L_9
\end{dmath}
and
\begin{align}   
\bar U(\epsilon)^{(8)} =&~    \frac{193583934848 L_1^{10}}{11023023375}-\frac{115326602912 \zeta _2 L_1^8}{734868225}-\frac{988689478016 \zeta _3
   L_1^7}{5144077575}\nonumber\\&+\frac{206442604544 L_4 L_1^6}{244956075}-\frac{4323741943184 \zeta _4 L_1^6}{244956075}+\frac{189258760192 L_5 L_1^5}{34993725}\nonumber\\&+\frac{99634322068672 \zeta _2
   \zeta _3 L_1^5}{1714692525}-\frac{31516238658832 \zeta _5 L_1^5}{244956075}-\frac{1823443023928 \zeta _3^2 L_1^4}{146973645}\nonumber\\&+32768 L_6 L_1^4-\frac{154731350528 L_4 \zeta _2
   L_1^4}{48991215}-\frac{1879223542881 \zeta _6 L_1^4}{4665830}\nonumber\\&+\frac{77932463360 \zeta _{-5,-1} L_1^4}{3266081}-\frac{84441751552 L_5 \zeta _2 L_1^3}{859495}+\frac{7659324416 L_4
   \zeta _3 L_1^3}{2332915}\nonumber\\&-\frac{124113413905792 \zeta _3 \zeta _4 L_1^3}{114312835}+\frac{13448253034912 \zeta _2 \zeta _5 L_1^3}{16330405}-\frac{9102340863776 \zeta _7
   L_1^3}{16330405}\nonumber\\&-\frac{139551801344 \zeta _{-5,1,1} L_1^3}{16330405}+\frac{338954846208 \zeta _{5,-1,-1} L_1^3}{16330405}-\frac{49152}{5} L_4^2 L_1^2\nonumber\\&+\frac{283482885319456 \zeta _2
   \zeta _3^2 L_1^2}{342938505}-294912 L_6 \zeta _2 L_1^2+\frac{127585738752 L_5 \zeta _3 L_1^2}{2332915}\nonumber\\&-\frac{979150630144 L_4 \zeta _4 L_1^2}{1256185}-\frac{35190890820656 \zeta _3
   \zeta _5 L_1^2}{16330405}+\frac{14236192309081 \zeta _8 L_1^2}{12032930}\nonumber\\&+\frac{17875425099264 \zeta _{-7,-1} L_1^2}{16330405}-\frac{51822569782272 \zeta _2 \zeta _{-5,-1}
   L_1^2}{114312835}\nonumber\\&+\frac{13597543296 \zeta _{5,3} L_1^2}{251237}-\frac{1244595191808 \zeta _{-5,-1,-1,-1} L_1^2}{16330405}\nonumber\\&+\frac{131915132928 \zeta _{-5,-1,1,1}
   L_1^2}{2332915}-\frac{84491033992672 \zeta _3^3 L_1}{342938505}\nonumber\\&-\frac{1116426125312 L_4 L_5 L_1}{16330405}-196608 L_7 \zeta _2 L_1+\frac{651122749440 L_6 \zeta _3
   L_1}{3266081}\nonumber\\&+\frac{23571268601856 L_4 \zeta _2 \zeta _3 L_1}{16330405}-\frac{23804318298624 L_5 \zeta _4 L_1}{16330405}\nonumber\\&-\frac{6115751323648 L_4 \zeta _5
   L_1}{2332915}+\frac{687025452705776 \zeta _4 \zeta _5 L_1}{114312835}\nonumber\\&-\frac{482730659185321 \zeta _3 \zeta _6 L_1}{228625670}+\frac{3284687600144 \zeta _2 \zeta _7
   L_1}{16330405}\nonumber\\&+\frac{291569715153149 \zeta _9 L_1}{48991215}+\frac{22123700403968 \zeta _3 \zeta _{-5,-1} L_1}{22862567}\nonumber\\&-\frac{194572778496 \zeta _{-7,1,1}
   L_1}{179455}+\frac{9141735161856 \zeta _2 \zeta _{-5,1,1} L_1}{16330405}\nonumber\\&+\frac{5249247401984 \zeta _2 \zeta _{5,-1,-1} L_1}{16330405}-\frac{9735062332416 \zeta _{7,-1,-1}
   L_1}{16330405}\nonumber\\&-\frac{98529325056 \zeta _{-5,-1,-1,-1,1} L_1}{2332915}+\frac{461763182592 \zeta _{-5,-1,-1,1,1} L_1}{16330405}\nonumber\\&-\frac{37101842432 \zeta _{-5,-1,1,-1,-1}
   L_1}{3266081}+\frac{227378528256 \zeta _{-5,-1,1,1,1} L_1}{1256185}\nonumber\\&-\frac{622331584512 L_5^2}{1256185}-\frac{228940688320 L_4 \zeta _3^2}{753711}-\frac{24032755897757 \zeta
   _5^2}{8793295}\nonumber\\&+786432 L_4 L_6-786432 L_{10}+\frac{334640510976 L_4^2 \zeta _2}{1256185}-638976 L_8 \zeta _2\nonumber\\&+\frac{26484736 L_7 \zeta _3}{35}+\frac{1883485807104 L_5 \zeta _2 \zeta
   _3}{8793295}-\frac{29574019653139 \zeta _3^2 \zeta _4}{158279310}\nonumber\\&-\frac{2364489123072 L_6 \zeta _4}{1256185}-\frac{1402421188992 L_5 \zeta _5}{1256185}-\frac{9041030424546 \zeta _2
   \zeta _3 \zeta _5}{8793295}\nonumber\\&-\frac{9533825728788 L_4 \zeta _6}{1256185}-\frac{7749622338496 \zeta _3 \zeta _7}{3768555}+\frac{513309731999841631 \zeta
   _{10}}{28138544000}\nonumber\\&-\frac{3294308047872 \zeta _{-9,-1}}{251237}+\frac{31484349153024 \zeta _2 \zeta _{-7,-1}}{8793295}-\frac{85147736064 L_4 \zeta
   _{-5,-1}}{179455}\nonumber\\&+\frac{23131188644304 \zeta _4 \zeta _{-5,-1}}{8793295}+\frac{2882611601582 \zeta _2 \zeta _{5,3}}{43966475}-\frac{899943270451 \zeta
   _{7,3}}{17586590}\nonumber\\&+\frac{1295698469376 \zeta _3 \zeta _{-5,1,1}}{8793295}-\frac{6327273143552 \zeta _3 \zeta _{5,-1,-1}}{8793295}\nonumber\\&+\frac{854235614208 \zeta
   _{-7,-1,-1,-1}}{1256185}+\frac{1070657270784 \zeta _{-7,-1,1,1}}{1256185}\nonumber\\&-\frac{377390456832 \zeta _2 \zeta _{-5,-1,-1,-1}}{8793295}+\frac{621206486016 \zeta _2 \zeta
   _{-5,-1,1,1}}{8793295}\nonumber\\&+\frac{35739672576 \zeta _{-5,1,-1,-1,-1,-1}}{1256185}+\frac{49152}{5} \zeta _{-5,1,-1,1,-1,-1}-\frac{98304}{5} \zeta _{-5,1,1,-1,-1,1}\nonumber\\&+\frac{39492169728 \zeta
   _{-5,1,1,-1,1,-1}}{1256185}-\frac{45016952832 \zeta _{-5,1,1,1,-1,-1}}{1256185}\nonumber\\&+\frac{9277956096 \zeta _{-5,1,1,1,1,1}}{1256185}
\end{align}

\bibliographystyle{JHEP}

\bibliography{biblio3}

\end{document}